\documentclass[journal,twoside,web]{IEEEtran}
\usepackage{cite}
\usepackage{amsmath,amssymb,amsfonts}
\usepackage{algorithmic}
\usepackage{romannum}
\usepackage{graphicx}
\usepackage{multirow}
\usepackage{pifont}
\usepackage{textcomp}
\usepackage{hyperref}

\usepackage{algorithm,algorithmic}
\makeatletter
\newcommand{\RNum}[1]{\uppercase\expandafter{\romannumeral #1\relax}}
\makeatother

\begin{document}
\title{Pseudo Label-Guided Data Fusion and Output Consistency for Semi-Supervised Medical Image Segmentation}
\author{Tao Wang, Yuanbin Chen, Xinlin Zhang, Yuanbo Zhou, Junlin Lan, Bizhe Bai, Tao Tan, Min Du, Qinquan Gao, Tong Tong
	\thanks{This work was supported by National Natural Science Foundation of China under Grant 62171133, in part by the Artificial Intelligence and Economy Integration Platform of Fujian Province, and the Fujian Health Commission under Grant 2022ZD01003.}
	\thanks{Tao Wang, Yuanbin Chen, Xinlin Zhang, Yuanbo Zhou, Junlin Lan, Min Du, Qinquan Gao and Tong Tong are with the college of physics and information engineering, University of Fuzhou University, Fuzhou 350108, China (e-mail: ortonwangtao@gmail.com; binycn904363330@gmail.com; xinlin1219@gmail.com; webbozhou@gmail.com; smurfslan@qq.com; dm$\_$dj90@163.com; gqinquan@imperial-vision.com; ttraveltong@imperial-vision.com).}
	\thanks{Tao Tan is with the Faculty of Applied Science, University of Macao Polytechnic University, Macao 999078 (e-mail: taotanjs@gmail.com).}
	\thanks{Bizhe Bai is with the University of Toronto University, Ontario M5S2E8, Canada (e-mail: bizhe.bai@outlook.com).} 
}

\maketitle

\begin{abstract}
	Supervised learning algorithms based on Convolutional Neural Networks have become the benchmark for medical image segmentation tasks, but their effectiveness heavily relies on a large amount of labeled data. However, annotating medical image datasets is a laborious and time-consuming process. Inspired by semi-supervised algorithms that use both labeled and unlabeled data for training, we propose the PLGDF framework, which builds upon the mean teacher network for segmenting medical images with less annotation. We propose a novel pseudo-label utilization scheme, which combines labeled and unlabeled data to augment the dataset effectively. Additionally, we enforce the consistency between different scales in the decoder module of the segmentation network and propose a loss function suitable for evaluating the consistency. Moreover, we incorporate a sharpening operation on the predicted results, further enhancing the accuracy of the segmentation.
	
	Extensive experiments on three publicly available datasets demonstrate that the PLGDF framework can largely improve performance by incorporating the unlabeled data. Meanwhile, our framework yields superior performance compared to six state-of-the-art semi-supervised learning methods. The codes of this study are available at https://github.com/ortonwang/PLGDF.
\end{abstract}


\section{Introduction}  % analysis \cite{8370732}
\IEEEPARstart{S}{egmentation} is a fundamental task in the field of medical image processing and analysis \cite{8370732}. Accurate image segmentation in clinical medicine provides valuable auxiliary information for clinicians, facilitating rapid, accurate, and efficient diagnostic decision-making \cite{anwar2018medical}. However, manual annotation of regions of interest is time-consuming and relies on the clinical expertise of physicians, resulting in a significant workload and potential errors \cite{sjoberg2013clinical}. 

With the rapid development of deep learning, Convolutional Neural Networks (CNN) and its variants have demonstrated powerful image processing capabilities in computer vision tasks. The introduction of Fully Convolutional Networks \cite{long2015fully} and U-Net \cite{ronneberger2015u} has greatly propelled the development of automated image segmentation \cite{dolz2018hyperdense}. Building upon these foundations, numerous studies have emerged to further improve the performance of segmentation algorithms \cite{9066969}\cite{9851919}\cite{10101800}.
For instance, Ning et al. proposed SMU-Net \cite{9551285}, which utilizes salient background representation to assist foreground segmentation by considering the texture information present in the background. Pang et al. introduced a novel two-stage framework named SpineParseNet for automated spine parsing in volumetric magnetic resonance images \cite{9201093}. Additionally, Chen et al. presented TransUNet \cite{chen2021transunet}, which combines CNN with Transformer \cite{NIPS2017_3f5ee243} for medical image segmentation, demonstrating outstanding performance. 

However, the success of these methods heavily relies on a large amount of pixel-level annotated data, which is only feasible through precise annotations by skilled medical professionals \cite{kohli2017medical}. This process is time-consuming and costly, limiting the applicability of supervised learning methods. 

To address this issue, researchers have proposed semi-supervised learning-based methods for medical image segmentation \cite{tarvainen2017mean}. 
Compared to supervised learning, semi-supervised learning can fully utilize the information contained in unlabeled data, thus improving the generalization capability and accuracy of the segmentation model \cite{10120761}.

One common approach in semi-supervised learning is the use of pseudo-label strategies. This method typically employs labeled data to train an initial model, which is then applied to unlabeled data to generate pseudo-labels. These pseudo-labels serve as approximate labels for the unlabeled data, thereby expanding the labeled dataset. Subsequently, the model is retrained using the expanded dataset to improve its robustness \cite{zheng2020cartilage} \cite{zheng2020semi} Qiu et al. introduced a Federated Semi-Supervised Learning \cite{10121665} approach to learn from distributed medical image domains, incorporating a federated pseudo-labeling strategy for unlabeled clients to mitigate the deficiency of annotations in unlabeled data. Bai et al. proposed an iterative learning method based on pseudo-labeling for cardiac MR image segmentation \cite{bai2017semi}. In this approach, pseudo-labels are refined using a Conditional Random Field, and the updated pseudo-labels are utilized for model updating. 

Another common approach in semi-supervised learning is the consistency-based method \cite{yu2019uncertainty}. This method aims to enhance the model's robustness by combining the consistency among unlabeled data. In the context of image segmentation tasks, consistency can be categorized into data-level consistency and model-level consistency. Data-level consistency requires the model to produce consistent predictions for different perturbations of the same image. For example, when introducing slight perturbations or applying different data augmentation techniques to the input image, the model should generate the same segmentation results. On the other hand, model-level consistency requires consistent segmentation results across different models for the same input.

Deep adversarial training \cite{9966841}\cite{CHEN2021118568} is also a commonly used method that leverages unlabeled data by employing a discriminator to align the distributions of labeled and unlabeled data. Wu et al. introduced MC-Net \cite{wu2022mutual}, which comprises a shared encoder and multiple slightly different decoders. The model incorporates statistical differences among the decoders to represent the model's uncertainty and enforce consistency constraints.

Furthermore, the mean-teacher model \cite{tarvainen2017mean} and its extensions \cite{li2020transformation}\cite{wang2022semi}\cite{xu2023ambiguity} have gained significant attention in semi-supervised medical image segmentation tasks. 
In the mean-teacher model, the parameter of the student network is guided by the teacher network during the training process. The model training involves minimizing the error between the teacher and student models. Additionally, other algorithms have also demonstrated outstanding performance in this field. 

Luo et al. proposed Uncertainty Rectification Pyramid Consistency URPC \cite{luo2022semi}, a novel framework with uncertainty rectified pyramid consistency regularization. This framework offers a straightforward and efficient method to enforce output consistency across various scales for unlabeled data. 

While the framework is simple and efficient, there is still room for further optimization in its performance, indicating the potential for further improvements in the performance of these methods. Therefore, this study also endeavors to explore some integrated strategies to enhance the performance of semi-supervised learning algorithms.
We propose a novel framework named Pseudo Label-Guided Data Fusion (PLGDF). The main contributions of this paper can be summarized as follows:
\begin{itemize}
	\item We introduce the PLGDF framework, a novel architecture built upon the mean teacher network, incorporating an innovative pseudo-label utilization scheme. This framework integrates consistency evaluation across various scales within the decoder module of the network.
	\item The mixing module combines both labeled and unlabeled data, enhancing the dataset's diversity. Additionally, the integrated sharpening operation further improves recognition accuracy.
	\item Experimental results on three publicly available datasets demonstrate the superiority of the proposed approach in semi-supervised medical image segmentation compared to six state-of-the-art models, setting new performance benchmarks.
\end{itemize}

\section{Relate Work}
\subsection{Medical image segmentation}
The advancement of deep learning has significantly enhanced the precision of semantic segmentation. Within medical image segmentation, U-Net and its extensions have become the benchmark methods for further research and practical applications. Building upon the foundation of U-Net, numerous high-performing algorithms have emerged, such as CE-Net\cite{8662594}, UNet++ \cite{8932614}, V-Net \cite{milletari2016v}, and the introduction of 3D U-Net \cite{10.1007/978-3-319-46723-8_49}, expanding the application of medical image segmentation into the realm of 3D medical images. Cao et al. proposed Swin-Unet \cite{10.1007/978-3-031-25066-8_9}, substituting convolutional blocks with Swin-Transformer \cite{Liu_2021_ICCV} blocks for enhanced feature extraction, while Wang et al. introduced O-Net \cite{wang2022net}, a deeper integration of CNN and Transformer, further improving algorithm performance. Additionally, other algorithms such as UNeXt \cite{10.1007/978-3-031-16443-9_3}, SpineParseNet \cite{9201093}, and SegFormer \cite{NEURIPS2021_64f1f27b} have contributed to the improvement of the model.
Although these methods have achieved success in medical image segmentation, they are predominantly constructed in a fully-supervised manner. Their performance is notably constrained by the scarcity of labeled samples available for training.
\begin{figure*}[!ht]
	\centering
	\includegraphics[width=1\textwidth]{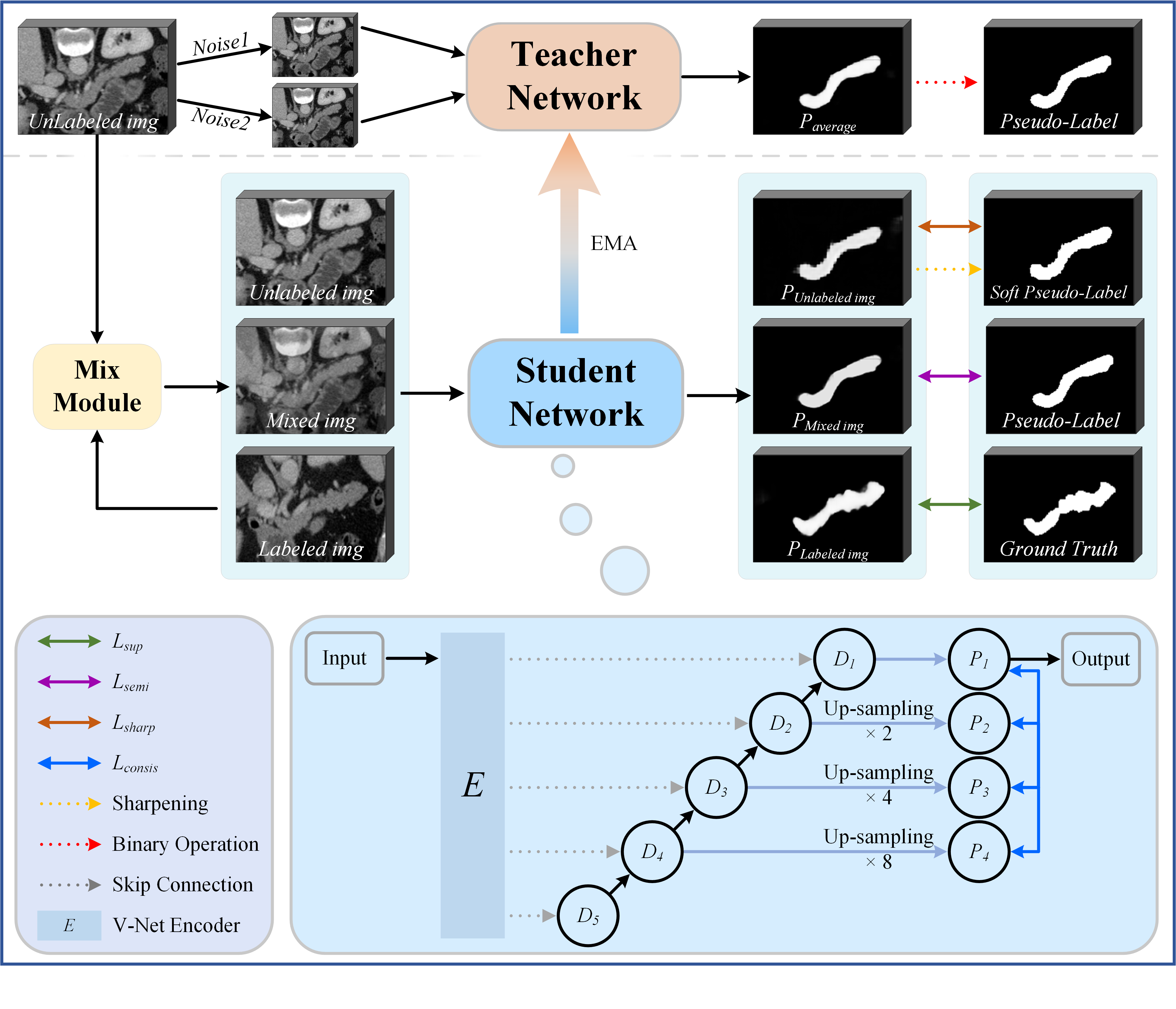}		
	\caption{The schematic diagram of the proposed PLGDF framework. The figure in the bottom right provides a detailed depiction of the student network. In the diagram, $D_{i}$ represents the decoder module of the V-Net backbone, and $P_{i}$ refers to the predictions obtained from different scales of the backbone. These predictions are unified to the same size through upsampling and convolution processes at various scales. Additionally, the Exponential Moving Average (EMA) represents that the teacher network implements parameter updates from the student network through the exponential moving average.}
	\label{figure1}
\end{figure*}
\subsection{Semi-Supervised Medical image segmentation}
To address the challenge of limited labeled data in medical image segmentation, researchers have proposed various semi-supervised learning methods. Currently, one of the most widely employed approaches involves extending the mean-teacher framework to different aspects. For example, Li et al. introduced TCSM \cite{li2020transformation} which leverages various perturbations on unlabeled data to train the network by enforcing consistency in predictions through regularization. Yu et al. presented Uncertainty-Aware Mean Teacher (UA-MT) \cite{uamt000}, which incorporates an uncertainty-aware scheme encouraging consistent predictions for the same input under different perturbations.
Chen et al. proposed an enhancement of the mean-teacher framework combined with adversarial networks to distinguish labeled and unlabeled data, showcasing outstanding performance \cite{CHEN2021118568}.
Furthermore, there is research based on task consistency. Luo et al. utilized a dual-task deep network for joint prediction of pixel segmentation maps and hierarchical representations of geometric objects, along with the introduction of dual-task consistency regularization[18]. 

Pseudo-label \cite{Chen_2021_CVPR} is another common semi-supervised learning framework, often involving the conversion of probability mappings into pseudo-labels using sharpening functions or fixed thresholds. Li et al proposed self-loop uncertainty, a pseudo-label application strategy where recurrent optimization of the neural network with a self-supervised task generates ground-truth labels for unlabeled images, augmenting the training set and enhancing segmentation accuracy. Rizve et al \cite{rizve2021defense} unified probability and uncertainty thresholds to select the most accurate pseudo-labels.
Luo et al presented URPC, which leverages data pyramid consistency and uncertainty rectification within a single model, based on the consistency of outputs at different scales for the same input, achieving excellent performance in semi-supervised learning for segmentation \cite{luo2022semi}.
Building upon previous attempts, we have combined pseudo label and pyramid Consistency with the mean teacher framework, to further improve the semi-supervised segmentation in medical images.
\section{Methods}
To provide a more comprehensive understanding of our proposed model, we begin by introducing the symbols utilized in our approach to address the problem of semi-supervised segmentation. Symbol definitions are as follows:\\
$X_{l}$: the labeled data, where each sample in $X_{l}$ is associated  with its corresponding Ground Truth, denoted as $GT$. \\
$X_{u}$: the unlabeled data, which consists of samples in the dataset that do not have corresponding $GT$. \\
$X_{mix}$: the data mixed from $X_{u}$ and $X_{l}$ through Mix Module during the training process. \\
$f_{\theta_{s}}(x)$: the generated probability map of input $x$, where $\theta_{s}$ denotes the parameters of the student network, and $f_{\theta_{t}}(x)$ means the generated probability map by the teacher network.   % 要不要介绍 theta 学生和教师

\begin{algorithm}[!ht]
	%\DontPrintSemicolon
	%	\begin{algorithm}
	\textbf{Input:} ($X_{l}$, $GT$) $\in$ $D_{l}$, ($X_{u}$) $\in$ $D_{u}$ \\
	\textbf{Output:} Final parameters of the resultant student model
	\hspace*{3.7em} $\theta_{s}$ and $\theta_{t}$ for teacher model \\
	%	\hspace*{0.5em}1: $f_{\theta_{s}}(x) = $ student network with parameters $\theta_{s}$ \\
	%	\hspace*{0.5em}2: $f_{\theta_{t}}(x) = $ teacher network with parameters $\theta_{t}$ \\
	\hspace*{0.5em}1: $BestVal() =$ function to select parameters of the model \\
	\hspace*{7.2em} with best validation performance \\
	\hspace*{0.5em}2: Other notations used are as described in the paper \\ \\
	\hspace*{0.5em}3:	\textbf{for $iter = 1, . . .\ , itermax$ do} \\
	\hspace*{0.5em}4:  \quad	$X_{u_{i}}=Random\ Noise^{i}(X_{u}), i \in(1,2)$ \\\\
	\hspace*{0.5em}5:  \quad	Acquire $f_{\theta _{t}}(X_{u_{i}}),i \in(1,2)$\\ 
	\hspace*{0.5em}6:  \quad	 $pseudo\ label=Argmax(\frac{1}{2}\textstyle \sum_{i=1}^{2}f_{\theta _{t}}(X_{u_{i}})$ \\					
	\hspace*{0.5em}7:  \quad	$X_{mix}=Mix\ Module(X_{u}, X_{l})$\\
	\hspace*{0.5em}8:  \quad	Generare $f_{\theta_{s}}(X_{l})$, $f_{\theta_{s}}(X_{mix})$, $f_{\theta_{s}}( X_{u})$\\
	\hspace*{0.5em}9:  \quad	$X = X_{u}\cup X_{mix}  $\\
	10:	\quad	$P_{1},\ P_{2},\ P_{3},\ P_{4} = f_{\theta_{s}}(X)$\\
	11:	\quad	$soft\ pseudo\ label = Sharpening(f_{\theta_{s}}(X_{u}))$ \\ \\
	12:	\quad Computing losses: \\
	13:	\quad $Loss_{sup} = L_{sup}(f_{\theta_{s}}(X_{l}),\ GT)  $ \\
	14:	\quad$Loss_{semi} =L_{semi}(f_{\theta_{s}}(X_{u}),\ pseudo\ label)  \\ 
	\hspace*{9em} + L_{semi}(f_{\theta_{s}}(X_{mix}),\ pseudo\ label)$ \\
	15:	\quad	$Loss_{sharp} = L_{sharp}(f_{\theta_{s}}(X_{u}),\ soft\ pseudo\ label)$ \\ 
	16:	\quad	$Loss_{consis} = L_{consis}(P_{1},\ P_{2},\ P_{3},\ P_{4})$	\\ 
	17:	\quad	$Loss_{total} = Loss_{sup}+Loss_{semi} + \\ 
	\hspace*{12.7em}      \lambda(Loss_{sharp}+Loss_{consis})$ \\ \\
	18:	\quad 	Minimize the $Loss_{total} $ for $\theta_{s}$\\
	19:	\quad 	$\theta_{t}\leftarrow \alpha\theta_{t} + (1-\alpha)\theta_{s}$ \\
	20:	\quad 	Save $\theta_{s}= BestVal (\theta_{s}$) \\
	21:	\textbf{end for} \\
	22: \textbf{return} $\theta_{s}$%, $\theta_{t}$
	
	\caption{Pseudo Label-Guided Data Fusion and Output Consistency for Semi-Supervised Medical Image Segmentation}\label{alg:alg1}
	
\end{algorithm}
\subsection{Overall architecture design} %In Algorithm 1, we present the pseudocode to illustrate the training procedure of the proposed
Figure \ref{figure1} illustrates the overall architecture of the PLGDF framework proposed in our study. In our methodology, we adopt the framework of the mean teacher model, where the teacher network implements parameter updates from the student network through exponential moving average (EMA). In Algorithm \ref{alg:alg1}, we present the pseudocode to illustrate the training procedure of the proposed framework.
To begin, we apply double random noise augmentation to the $X_{u}$. The teacher network processes the augmented data and obtains the average results, which are subsequently binarized to generate the pseudo-label corresponding to the $X_{u}$. Next, we utilize the Mix Module to augment the $X_{u}$ with $X_{l}$, resulting in the $X_{mix}$ which is displayed as $Mixed$ $ img$ in the Figure \ref{figure1}. Finally, we concatenate the $X_{u}$, $X_{l}$ and $X_{mix}$, which are then processed by the student network to generate the corresponding prediction: $f_{\theta_{s}}(X_{u})$, $f_{\theta_{s}}(X_{l})$, and $f_{\theta_{s}}(X_{mix})$. Next, we further refine $f_{\theta_{s}}(X_{u})$ by applying a sharpening process to obtain soft pseudo-labels.

To facilitate the training of the model, we evaluate $L_{sup}$ based on $f_{\theta_{s}}(X_{l})$ and $GT$, $L_{semi}$ based on $f_{\theta_{s}}(X_{mix})$ and pseudo labels, $L_{sharp}$ based on $f_{\theta_{s}}(X_{u})$ and soft pseudo-labels. Additionally, we introduced multi-scale outputs for the backbone model, and a multi-scale consistency evaluation module is incorporated to assess the consistency among outputs from different scales, which is shown as $L_{consis}$.
In the subsequent subsections, we will provide a detailed explanation of each module.

\subsection{Mix Module}
To enhance the model's effectiveness and improve its robustness with limited data, we introduce a data augmentation technique inspired by the widely employed Mix-Up method in vision tasks \cite{zhang2017mixup}. We perform a data mixing process by combining $X_{l}$ and $X_{u}$, resulting in a more diverse training dataset. During the data mixing process, we randomly select two sets of samples and linearly interpolate their features by a certain proportion, generating new samples with blended characteristics for model training. For a pair of samples $X_{u_{1}}$ and $X_{l_{1}}$, this process can be mathematically represented as follows:
\begin{align}
\lambda = Random(Beta(\alpha ,\alpha )),\\
\lambda' = max(\lambda,1-\lambda),\\
X_{u_{1}}' = \lambda'X_{u_{1}} + (1 - \lambda')X_{l_{1}}
\end{align}
\begin{figure}[!h]
	\centering
	\includegraphics[width=0.47\textwidth]{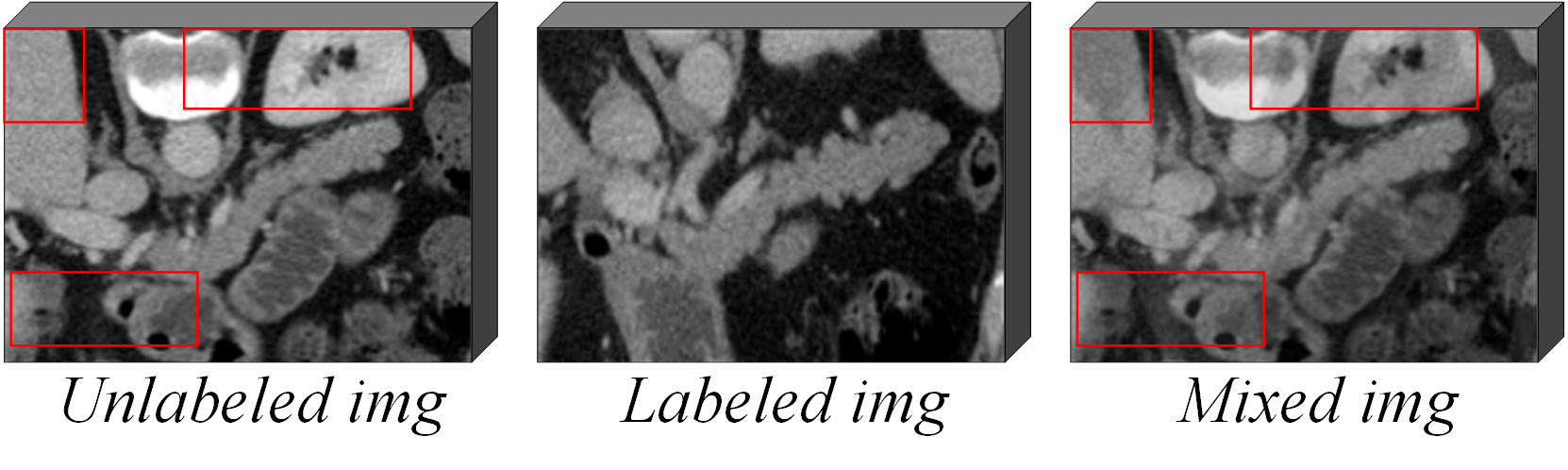}		
	\caption{The Renderings of Mix-Up.}
	\label{figure2}
\end{figure}

where $\alpha$ is a randomly generated hyperparameter. The effect of image mixing is shown in Figure \ref{figure2}. The mix is predominantly based on an unlabeled image, with a labeled image providing additional mixing. After the blending operation, certain changes occur in the information of the unlabeled image, especially within the confines of the red rectangular box, a more pronounced effect is showcased. However, these changes do not significantly alter the overall semantic information of the image. Therefore, we use the pseudo-label corresponding to the unlabeled image as the label for the mixed data.

\subsection{Pseudo label sharpening}
\begin{figure}[!h]
	\centering
	\includegraphics[width=0.47\textwidth]{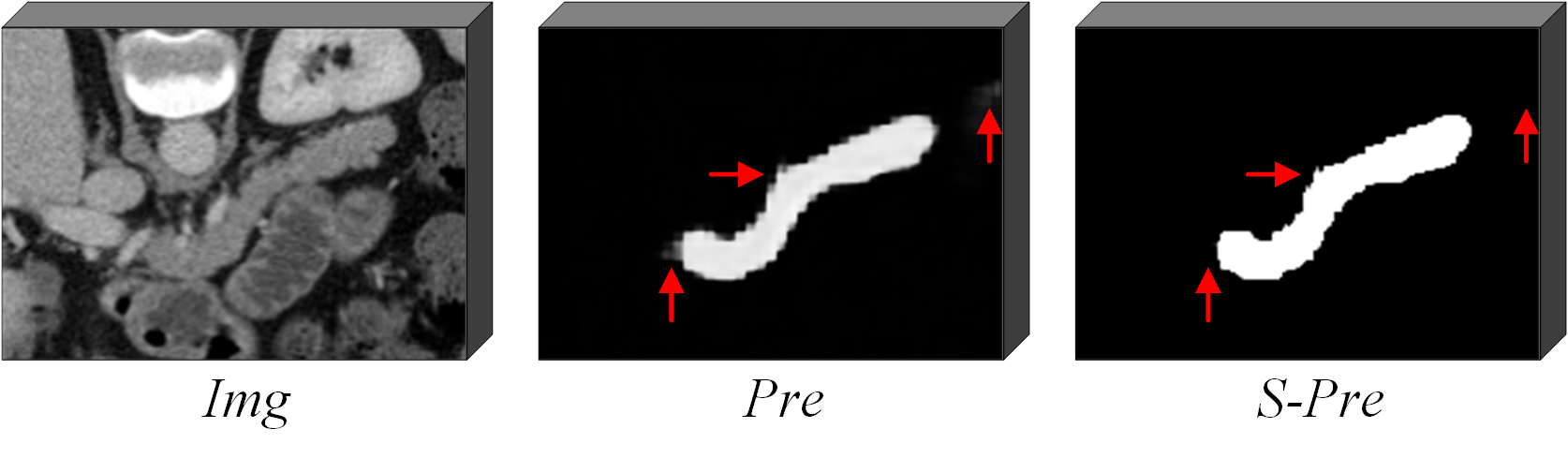}		
	\caption{Demonstration of the sharpening operation. The $Pre$ represents the prediction of the image,  the $S-Pre$ represents the results obtained after applying the sharpening operation to the $Pre$.}
	\label{figure3}
\end{figure}

Considering the efficacy of consistency in utilizing unlabeled data, we employ a sharpening function \cite{xie2020unsupervised} to transform the $f_{\theta_{s}}(X_{u})$ into soft pseudo-labels, which is the result obtained from the $X_{u}$ through the student network. The formula of the sharpening function is as follows:
\begin{equation}
f^{*}_{\theta_{s}}(X_{u}) = \frac{f_{\theta_{s}}(X_{u})^{\frac{1}{T} } }{f_{\theta_{s}}(X_{u}))^{\frac{1}{T} } + (1-f_{\theta_{s}}(X_{u}))^{\frac{1}{T} } } 
\end{equation}
Where $T$ is a hyperparameter used to control the sharpening temperature. 
Figure \ref{figure3} illustrates that the sharpening operation enhances the clarity and accuracy of segmentation boundaries, reducing blurriness and ambiguous edges, especially in the region indicated by the red arrow. 
This improvement enables a more effective capture of target boundaries and subtle structures. 
%This improvement contributes to a better performance of the model in areas with rich image details, enabling a more effective capture of target boundaries and subtle structures. 
Subsequently, we compute the consistency loss based on  $f_{\theta_{s}}(X_{u})$ and $f^{*}_{\theta_{s}}(X_{u})$. Under the supervision of soft pseudo-labels, the model learns to generate low-entropy results as a means of minimizing entropy. We denote this loss as $L_{sharp}$, with its evaluation formula expressed as follows:
\begin{equation}
L_{sharp} = \frac{1}{N}\sum_{i=1}^{N} ||f_{\theta_{s}}(X_{u}) - f^{*}_{\theta_{s}}(X_{u})||^{2} 
\end{equation}
Where $N$ is the total number of pixels.
% 上面的这个说明可能要改
\subsection{Consistency Across Multi-Scale} 
In the decoder part of the backbone network, convolution followed by upsampling operations is commonly utilized. Therefore, during the training phase, we can utilize the intermediate features at different scales within the decoder module and standardize their sizes by applying 2$\times$, 4$\times$, and 8$\times$ upsampling along with convolutional operations. Our objective is to enforce the consistency among outputs of different scales. 
%The specific evaluation rules for multi-scale consistency are described as follows:  %$\widehat{P}$
First, we calculate the average value $\widehat{P}$ based on the multi-scale $P_{1}, P_{2}, P_{3}$, and $P_{4}$ as shown in Figure \ref{figure1}. Subsequently, we assess the consistency between each scale and the average value $\widehat{P}$ using $L_{consis}$. %The specific evaluation formula is presented in equation (6), (7) and (8).
%However, relying solely on cosine distance may not meet our requirements enough. Inspired by $L_{urc}$ proposed in \cite{luo2022semi}, we propose $L_{consis}$ to mitigate errors caused by outlier pixels that significantly deviate from the average value due to insufficient labeled data. 
The evaluation formula is presented in equation (6), (7), and (8).
\begin{equation}
\widehat{P} =\frac{1}{4} \sum_{s=1}^{4}P_{s}  
\end{equation}
\begin{equation}
D_{s}^{i}  = \sum_{j=0}^{C-1}P_{s}^{ i,j } \cdot \log\frac{P_{s}^{i,j}}{\widehat{P}^{i,j }}  
\end{equation}
\begin{equation}
L_{consis} =\frac{1}{n} \sum_{s=1}^{n} \frac{ {\textstyle \sum_{i=1}^{N}} {||P_{s}^{i}  -\widehat{P} ^{i}||_{2}\cdot e^{-D_{s}^{i}}    } }{\sum_{i=1}^{N} e^{-D_{s}^{i}}}  + \sum_{i=1}^{N}D_{s} ^{i}
\end{equation}
%where $D_{s}^{i}$ is the KL-Divergence between $\widehat{P}$ and $P_{s}$ at pixel $i$,
where $C$ means the class number for the segmentation task, the $n$ represents the number of multi-scale outputs, and the $N$ is the total number of pixels.

\subsection{Loss Function}
The proposed PLGDF framework aims to learn from both labeled and unlabeled data by minimizing the following composite objective function:
\begin{equation}
L_{total} = L_{sup}+L_{semi}+ \lambda \cdot  ( L_{sharp} + L_{consis})
\end{equation}
the expressions for $L_{sharp}$ and $L_{consis}$ are already illustrated in equations (5) and (8), respectively. Regarding $L_{sup}$ and $L_{semi}$, we adopt the combination of Cross-Entropy Loss and Dice Loss \cite{milletari2016v}, which is commonly used in medical image segmentation. The formula for Dice Loss is presented below:
\begin{equation}
Dice \ Loss=1-\frac{2 * \sum_{i=1}^{N} p_{i} *g_{i} }{\sum_{i=1}^{N}p_{i} ^{2}  + \sum_{i=1}^{n}g_{i} ^{2} }  
\end{equation}
where $p_{i}$ is the value of the pixel $i$ predicted by the model, and $g_{i}$ is the value of the pixel $i$ of the ground truth.
We introduce $\lambda(t)$, a widely used time-dependent Gaussian warming-up function \cite{laine2016temporal}, to control the proportion between supervised and unsupervised loss at different training stages. Its specific formula is defined as follows:
\begin{equation}
\lambda(t) = w\cdot e^{(-5(1-\frac{t}{t_{max} } )^{2} )}
\end{equation}
where $w$ represents the final regulation weight,  $t$ represents the current training step and $t_{max}$ denotes the maximum training step.

%\begin{equation}
%$\lambda(t) = w\cdot e^{(-5(1-\frac{t}{t_{max} } )^{2} )}$
%Pturb\ seg=\frac{1}{N} \sum_{j=1}^{N} Pturb\ cls_{j}
%%	Pturb_{seg}=\frac{1}{N} \sum_{j=1}^{N}\frac{1}{4}\sum_{i=1}^{4}(P_{img\,j\,i}-P_{average\,j})^{2}  
%%	\end{gathered}
%\end{equation}

\section{Experiments and Results}

\subsection{Dataset}
In this paper, we evaluated the proposed PLGDF method and compared it with six previous works on three publicly available datasets: the Pancreas-CT dataset, the LA dataset and the BraTS2019 dataset. All of the datasets are associated with 3D segmentation tasks.
\subsubsection{Pancreas-CT}
The Pancreas-CT dataset \cite{clark2013cancer} consists of 82 3D abdominal contrast-enhanced CT scans, acquired from a cohort of 53 male and 27 female subjects. The CT scans were obtained with resolutions of 512$\times$512 pixels and varying pixel sizes. The slice thickness ranged from 1.5 to 2.5 mm. For this study, we randomly selected 60 images for training and 20 images for testing, following a standard data splitting protocol commonly used in similar studies \cite{wu2022mutual}. %这里插入跟哪个数据集一样
To ensure consistency and comparability of voxel values, we applied a clipping operation, limiting the values to the range of -125 to 275 Hounsfield Units (HU) \cite{zhou2019prior}. Additionally, we performed data resampling to achieve an isotropic resolution of 1.0$\times$1.0$\times$1.0 mm.
\subsubsection{LA}
The LA dataset \cite{xiong2021global}, which serves as the benchmark dataset for the 2018 Atrial Segmentation Challenge, comprises 100 gadolinium-enhanced MR imaging scans for training, with a resolution of 0.625$\times$0.625$\times$0.625 mm. As the testing set of LA lacks publicly available annotations, we allocated 80 samples for training and reserved the remaining 20 samples for validation following \cite{wu2022mutual}. Subsequently, we evaluated the performance of our model and other methods on the same validation set to ensure fair comparisons.
\subsubsection{BraTS2019}
The publicly available BraTS2019 dataset\cite{menze2014multimodal} comprises scans obtained from 335 patients diagnosed with glioma. This dataset encompasses T1, T2, T1 contrast-enhanced, and FLAIR sequences, along with corresponding tumor segmentations annotated by expert radiologists. In this study, we focused on using the FLAIR modality for segmentation on the dataset. We conducted a random split, allocating 250 scans for training, 25 scans for validation, and 60 scans for testing following \cite{luo2022semi}.

%\label{tabel2}

% pa 的表
\begin{table*}[!h]
	\centering
	\caption{Quantitative comparison with six state-of-the-art methods on the Pancreas-CT dataset. }
	\label{table1}
	\begin{tabular}{clcclcccc}
		\hline
		\multirow{2}{*}{Method} &  & \multicolumn{2}{c}{Scans used}                        &  & \multicolumn{4}{c}{Metrics}                                                                   \\ \cline{3-4} \cline{6-9} 
		&  & Labeled                   & Unlabeled                 &  & Dice(\%)$\uparrow$ & Jaccard(\%)$\uparrow$ & 95HD(voxel)$\downarrow$ & ASD(voxel)$\downarrow$ \\ \hline
		V-Net                   &  & 3(5\%)                   & 0                         &  & 29.32              & 19.61                 & 43.67                   & 15.42                  \\
		V-Net                   &  & 6(10\%)                   & 0                         &  & 54.94              & 40.87                 & 47.48                   & 17.43                  \\
		V-Net                   &  & 12(20\%)                  & 0                         &  & 71.52              & 57.68                 & 18.12                   & 5.41                   \\
		V-Net                   &  & 62(100\%)                 & 0                         &  & 83.76              & 72.48                 & 4.46                    & 1.07                   \\ \hline
		UA-MT (MICCAI'19)       &  & \multirow{7}{*}{3(5\%)}   & \multirow{7}{*}{59(95\%)} &  & 43.15              & 29.07                 & 51.96                   & 20.00                  \\
		SASSNet (MICCAI'20)      &  &                           &                           &  & 41.48              & 27.98                 & 47.48                   & 18.36                  \\
		DTC (AAAI'21)            &  &                           &                           &  & 47.57              & 33.41                 & 44.17                   & 15.31                  \\
		URPC (MIA'2022 )         &  &                           &                           &  & 45.94              & 34.14                 & 48.80                   & 23.03                  \\
		SS-Net (MICCAI'22)       &  &                           &                           &  & 41.39              & 27.65                 & 52.12                   & 19.37                  \\
		MC-Net+ (MIA'2022 )      &  &                           &                           &  & 32.45              & 21.22                 & 58.57                   & 24.84                  \\
		Ours                    &  &                           &                           &  & \textbf{74.69}     & \textbf{60.00}        & \textbf{8.19}           & \textbf{1.74}          \\ \hline
		UA-MT (MICCAI'19)       &  & \multirow{7}{*}{6(10\%)}  & \multirow{7}{*}{56(90\%)} &  & 66.44              & 52.02                 & 17.04                   & 3.03                   \\
		SASSNet (MICCAI'20)     &  &                           &                           &  & 68.97              & 54.29                 & 18.83                   & 1.96                   \\
		DTC (AAAI'21)           &  &                           &                           &  & 66.58              & 51.79                 & 15.46                   & 4.16                   \\
		URPC (MIA'2022 )        &  &                           &                           &  & 73.53              & 59.44                 & 22.57                   & 7.85                   \\
		SS-Net (MICCAI'22)       &  &                           &                           &  & 73.44              & 58.82                 & 12.56                   & 2.91                   \\
		MC-Net+ (MIA'2022 )     &  &                           &                           &  & 70.00              & 55.66                 & 16.03                   & 3.87                   \\
		Ours                    &  &                           &                           &  & \textbf{80.90}     & \textbf{68.40}        & \textbf{6.02}           & \textbf{1.59}          \\ \hline
		UA-MT (MICCAI'19)       &  & \multirow{7}{*}{12(20\%)} & \multirow{7}{*}{50(80\%)} &  & 76.10              & 62.62                 & 10.84                   & 2.43                   \\
		SASSNet (MICCAI'20)     &  &                           &                           &  & 76.39              & 63.17                 & 11.06                   & 1.42                   \\
		DTC (AAAI'21)           &  &                           &                           &  & 78.27              & 64.75                 & 8.36                    & 2.25                   \\
		URPC (MIA'2022 )        &  &                           &                           &  & 80.02              & 67.30                 & 8.54                    & 1.98                   \\
		SS-Net (MICCAI'22)       &  &                           &                           &  & 78.68              & 65.96                 & 9.74                    & 1.91                   \\
		MC-Net+ (MIA'2022 )     &  &                           &                           &  & 79.37              & 66.83                 & 8.52                    & 1.72                   \\
		Ours                    &  &                           &                           &  & \textbf{82.76}     & \textbf{70.89}        & \textbf{4.85}           & \textbf{1.33}          \\ \hline
	\end{tabular}
\end{table*}

\subsection{Implementation Details}
%Following \cite{luo2021semi}\cite{wu2022mutual}\cite{li2020shape},  we conducted a cropping operation on the 3D samples according to the ground truth. For the LA or Pancreas-CT datasets, we applied enlarged margins of [10$\sim$20, 10$\sim$20, 5$\sim$10] or [25, 25, 0] voxels, respectively. Afterward, the scans were normalized to zero mean and unit variance to facilitate training. 
During the training process, we randomly extracted 3D patches from the preprocessed data. For the LA dataset, the patch size was set to 112 $\times$ 112 $\times$ 80, while for the Pancreas-CT and BraTS2019 datasets, the patch size was 96 $\times$ 96 $\times$ 96.
%After that, we applied 2D rotation and flip operations as data augmentation on the LA dataset. 
For all three datasets, we set the batch size to 4, where each batch consisted of two labeled patches and two unlabeled patches. The backbone network employed in our study is the V-Net \cite{milletari2016v}. Additionally, we made modifications to the network to generate multi-scale outputs, and the scales $n$ we used for evaluating the multi-scale consistency is set to 4. We trained our PLGDF model for 15$k$ iterations for Pancreas-CT and LA datasets and 30$k$ iterations for the BraTS2019 dataset, following the methodology described in \cite{wu2022mutual} and \cite{luo2022semi}.

During the testing phase, we employed a sliding window approach with a fixed stride to extract patches. Specifically, on the LA dataset, we utilized a sliding window of size 112 $\times$ 112 $\times$ 80 with a stride of 18 $\times$ 18 $\times$ 4. On the Pancreas-CT and BraTS2019 datasets, we used a sliding window with a size of 96 $\times$ 96 $\times$ 96 and a stride of 16 $\times$ 16 $\times$ 16. Subsequently, we reconstructed the patch-based predictions to obtain the final results for the entire volume.

In our training process, we employed the SGD optimizer with a momentum 0.9 and weight decay set to 1e-4. The learning rate was set to 1e-2 and the hyperparameter $T$ was set to 1e-1. In this study, we trained the network using 10$\%$ and 20$\%$ of the data on three representative semi-supervised datasets, following the data partitioning methods as described in \cite{yu2019uncertainty}\cite{li2020shape}\cite{luo2021semi}.
Our framework was implemented in PyTorch 1.12.0, utilizing an Nvidia RTX 3090 GPU with 24GB of memory.
For quantitative evaluation, we employed four metrics: Dice, Jaccard, the average surface distance (ASD), and the 95$\%$ Hausdorff Distance (95HD). During the training phase, as our model incorporates multi-scale outputs, we utilized the four-scale output within the student model and exclusively employed the highest-scale output within the teacher model. Similarly, during the inference phase of the network, we solely utilized the output from the highest scale, denoted as $P_{1}$ in Figure \ref{figure1}. Consequently, at this juncture, our backbone network is equivalent to the V-Net. The $P_{2}$, $P_{3}$, and $P_{4}$ were exclusively utilized within the student models during the training process.

\subsection{Comparison with Other Semi-supervised Methods:}
We compared our proposed framework with six state-of-the-art semi-supervised segmentation methods, including UA-MT \cite{uamt000}, Shape-aware Adversarial Network (SASSNet) \cite{li2020shape}, Dual-task Consistency Framework (DTC) \cite{luo2021semi}, Uncertainty Rectified Pyramid Consistency (URPC) \cite{luo2022semi}, SS-Net \cite{10.1007/978-3-031-16443-9_4} and Mutual Consistency Network (MC-Net+) \cite{wu2022mutual}.
Note that we utilized the official codes and results of UA-MT, SASSNet, URPC, DTC, SS-Net, and MC-Net, along with their publicly available data preprocessing schemes. We used the results in MC-Net as our benchmark.
\subsubsection{Results on the Pancreas-CT dataset}
\begin{figure*}[!h]
	\centering
	\includegraphics[width=1.00\textwidth]{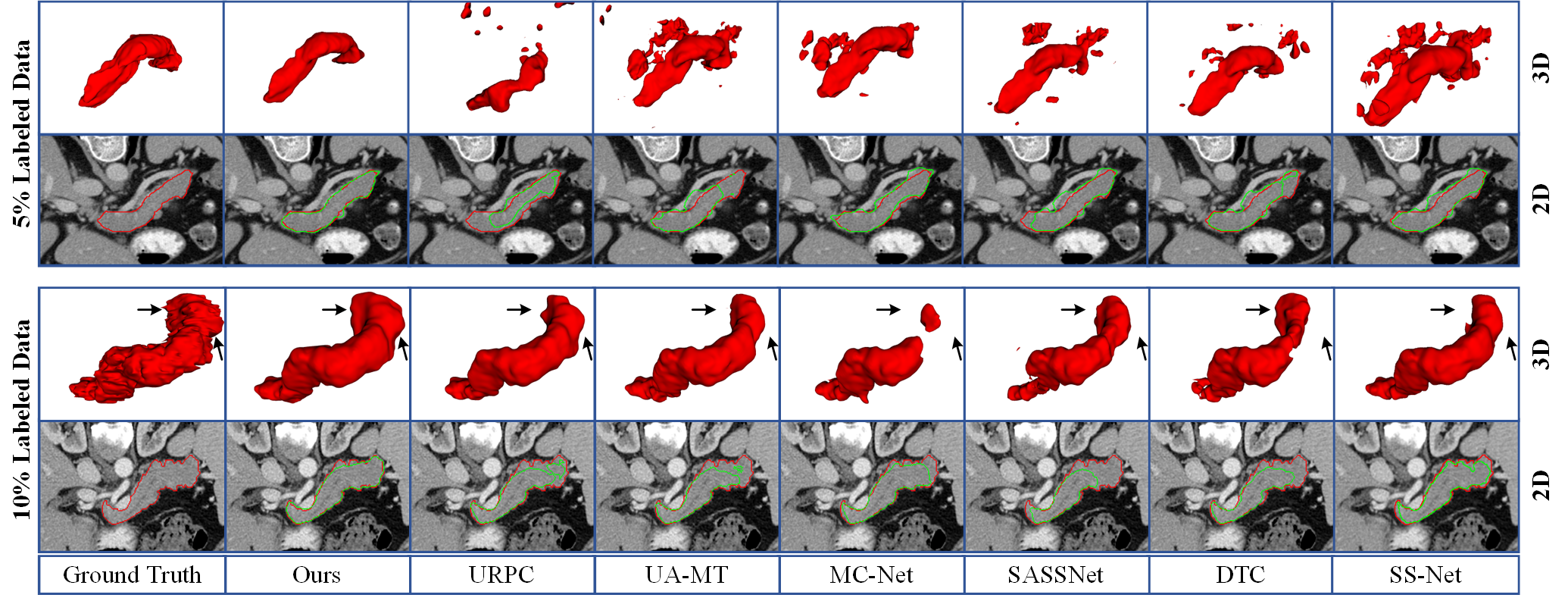}		
	\caption{2D and 3D Visualization with other methods on the Pancreas-CT dataset under 5$\%$ labeled data and 10$\%$ labeled data. The red lines denote the boundary of ground truth and the green lines denote the boundary of predictions.}
	\label{figure4}
\end{figure*}

% LA的表格
\begin{table*}[!h]
	\centering
	\caption{Quantitative comparison with six state-of-the-art methods on the LA dataset. }
	\label{table2}
	\begin{tabular}{ccccccccc}
		\hline
		\multirow{2}{*}{Method} &  & \multicolumn{2}{c}{Scans used}                        &  & \multicolumn{4}{c}{Metrics}                                                                   \\ \cline{3-4} \cline{6-9} 
		&  & Labeled                   & Unlabeled                 &  & Dice(\%)$\uparrow$ & Jaccard(\%)$\uparrow$ & 95HD(voxel)$\downarrow$ & ASD(voxel)$\downarrow$ \\ \hline
		V-Net                   &  & 4(5\%)                    & 0                         &  & 52.55              & 39.60                 & 47.05                   & 9.87                   \\
		V-Net                   &  & 8(10\%)                   & 0                         &  & 78.57              & 66.96                 & 21.20                   & 6.07                   \\
		V-Net                   &  & 16(20\%)                  & 0                         &  & 86.96              & 77.31                 & 11.85                   & 3.22                   \\
		V-Net                   &  & 80(100\%)                 & 0                         &  & 91.62              & 84.60                 & 5.40                    & 1.64                   \\ \hline
		UA-MT (MICCAI'19)       &  & \multirow{7}{*}{4(5\%)}   & \multirow{7}{*}{76(95\%)} &  & 82.26              & 70.98                 & 13.71                   & 3.82                   \\
		SASSNet (MICCAI'20)      &  &                           &                           &  & 81.60              & 69.63                 & 16.60                   & 3.58                   \\
		DTC (AAAI'21)            &  &                           &                           &  & 81.25              & 69.33                 & 14.90                   & 3.99                   \\
		URPC (MIA'2022 )         &  &                           &                           &  & 82.48              & 71.35                 & 14.65                   & 3.65                   \\
		SS-Net (MICCAI'22)       &  &                           &                           &  & 86.33              & 76.15                 & 9.97           & 2.31          \\
		MC-Net+ (MIA'2022 )      &  &                           &                           &  & 82.07              & 70.38                 & 20.49                   & 5.72                   \\
		Ours                    &  &                           &                           &  & \textbf{89.22}     & \textbf{80.62}        & \textbf{7.90}                   & \textbf{1.71}                   \\ \hline
		UA-MT (MICCAI'19)       &  & \multirow{7}{*}{8(10\%)}  & \multirow{7}{*}{72(90\%)} &  & 86.28              & 76.11                 & 18.71                   & 4.63                   \\
		SASSNet (MICCAI'20)     &  &                           &                           &  & 86.81              & 76.92                 & 12.54                   & 2.55                   \\
		DTC (AAAI'21)           &  &                           &                           &  & 87.51              & 78.17                 & 8.23                    & 2.36                   \\
		URPC (MIA'2022 )        &  &                           &                           &  & 85.01              & 74.36                 & 15.37                   & 3.96                   \\
		SS-Net (MICCAI'22)       &  &                           &                           &  & 88.43              & 79.43                 & 7.95                    & 2.55                   \\
		MC-Net+ (MIA'2022 )     &  &                           &                           &  & 88.96              & 80.25                 & 7.93                    & 1.86                   \\
		Ours                    &  &                           &                           &  & \textbf{89.8}      & \textbf{81.58}        & \textbf{7.14}           & \textbf{1.74}          \\ \hline
		UA-MT (MICCAI'19)       &  & \multirow{7}{*}{16(20\%)} & \multirow{7}{*}{64(80\%)} &  & 88.74              & 79.94                 & 8.39                    & 2.32                   \\
		SASSNet (MICCAI'20)     &  &                           &                           &  & 89.27              & 80.82                 & 8.83                    & 3.13                   \\
		DTC (AAAI'21)           &  &                           &                           &  & 89.42              & 80.98                 & 7.32                    & 2.10                   \\
		URPC (MIA'2022 )        &  &                           &                           &  & 88.74              & 79.93                 & 12.73                   & 3.66                   \\
		SS-Net (MICCAI'22)       &  &                           &                           &  & 89.86              & 81.70                 & 7.01                    & 1.87                   \\
		MC-Net+ (MIA'2022 )     &  &                           &                           &  & 91.07              & 83.67                 & 5.84                    & \textbf{1.67}          \\
		Ours                    &  &                           &                           &  & \textbf{91.34}     & \textbf{83.94}        & \textbf{5.66}           & 1.76                   \\ \hline
	\end{tabular}
\end{table*}

\begin{figure*}[!h]
	\centering
	\includegraphics[width=1.00\textwidth]{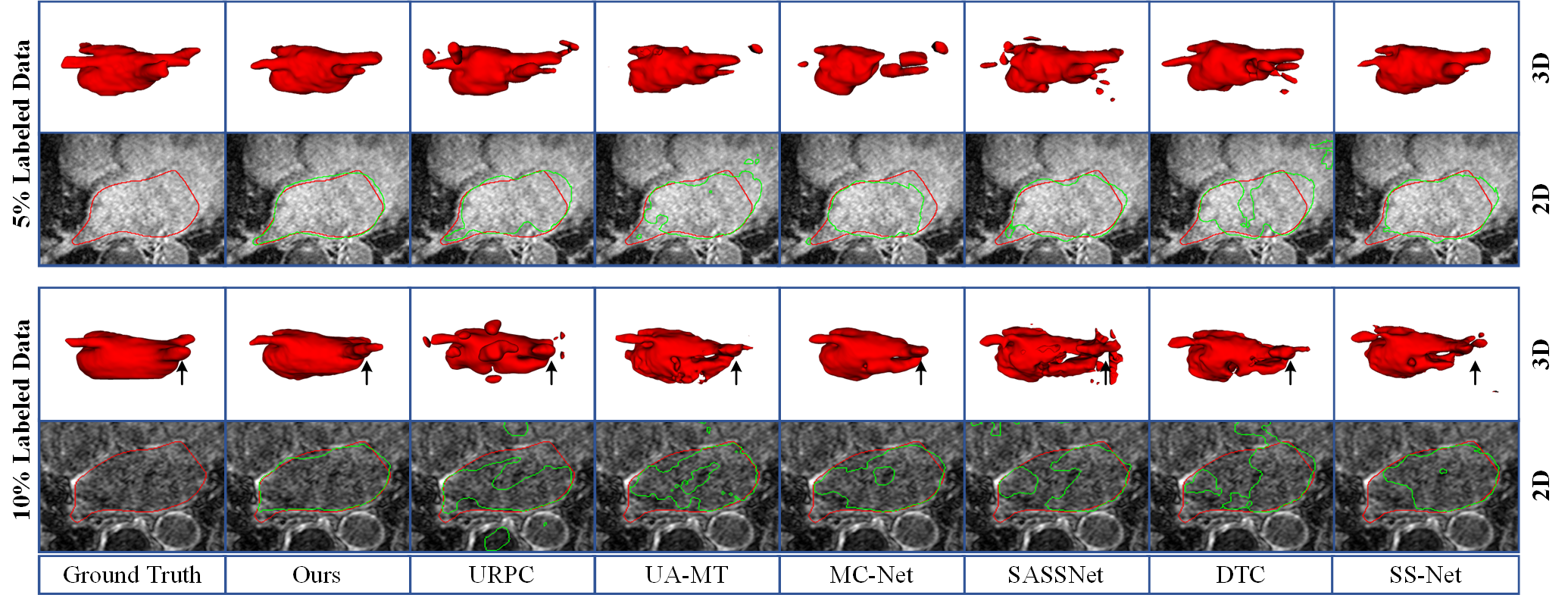}		
	\caption{2D and 3D Visualization with other methods on the LA dataset under 5$\%$ labeled data and 10$\%$ labeled data. The red lines denote the boundary of ground truth and the green lines denote the boundary of predictions.}
	\label{figure5}
\end{figure*}

Table \ref{table1} shows the quantitative comparison of our model and six semi-supervised methods on the Pancreas-CT dataset, along with the results of the V-Net model trained with 5$\%$,10$\%$, 20$\%$, and 100$\%$ labeled data for supervised learning. The experimental results indicate significant improvements in our proposed method over six compared state-of-the-art (SOTA) models across four evaluation metrics: Dice, Jaccard, 95HD, and ASD. 
It is evident from the table that our result stands out prominently, particularly when only 5$\%$ or 10$\%$ of the data is labeled. With 5$\%$ labeled data, we achieved a Dice score of 74.69$\%$, along with an 95HD of 8.19 and an ASD of 1.74. These metrics not only surpass the accuracy obtained by the six compared SOTA models based on equivalent data annotation but also outperform the accuracy of these models at 10$\%$ data annotation. Similarly, at 10$\%$ data annotation, the accuracy of our method exceeds those compared models trained with 20$\%$ labeled data annotation.

Figure \ref{figure4} displays the visual results of image segmentation on the Pancreas-CT dataset when trained with 5$\%$ and 10$\%$ labeled data. The visualizations are presented in both 2D and 3D perspectives. It is apparent that when utilizing only 5$\%$ labeling, other compared models fail to segment an approximate pancreas, while our model achieves segmentation results relatively closer to the Ground Truth (GT). Similarly, at 10$\%$ data annotation, our algorithm's predictions are more accurate. In the 2D visual representation, the comparative method exhibits a higher rate of false negatives, whereas our model achieves more precise identification. This further substantiates the effectiveness of the model we proposed.

% Brats 的表格
\begin{table*}[!h]
	\centering
	
	\caption{Quantitative comparison with six state-of-the-art methods on the BraTS2019 dataset.}
	\label{table3}
	\begin{tabular}{ccccccccc}
		\hline
		\multirow{2}{*}{Method} &  & \multicolumn{2}{c}{Scans used}                         &  & \multicolumn{4}{c}{Metrics}                                                                   \\ \cline{3-4} \cline{6-9} 
		&  & Labeled                   & Unlabeled                  &  & Dice(\%)$\uparrow$ & Jaccard(\%)$\uparrow$ & 95HD(voxel)$\downarrow$ & ASD(voxel)$\downarrow$ \\ \hline
		V-Net                   &  & 12(5\%)                   & 0                          &  & 74.28              & 64.42                 & 13.44                   & 2.60                   \\
		V-Net                   &  & 25(10\%)                  & 0                          &  & 78.67              & 68.75                 & 10.44                   & 2.23                   \\
		V-Net                   &  & 50(20\%)                  & 0                          &  & 80.59              & 71.13                 & 8.95                    & 2.03                   \\
		V-Net                   &  & 250(All)                  & 0                          &  & 88.58              & 80.34                 & 6.19                    & 1.36                   \\ \hline
		UA-MT (MICCAI'19)       &  & \multirow{7}{*}{12(5\%)}  & \multirow{7}{*}{238(95\%)} &  & 80.31              & 70.43                 & 10.65                   & \textbf{2.12}          \\
		SASSNet (MICCAI'20)      &  &                           &                            &  & 76.17              & 66.43                 & 13.09                   & 3.32                   \\
		DTC (AAAI'21)            &  &                           &                            &  & 74.21              & 64.89                 & 13.54                   & 3.16                   \\
		URPC (MIA'2022 )         &  &                           &                            &  & 78.74              & 68.2                  & 17.43                   & 4.51                   \\
		SS-Net (MICCAI'22)       &  &                           &                            &  & 78.03              & 68.11                 & 13.7           & 2.76                   \\
		MC-Net+ (MIA'2022 )      &  &                           &                            &  & 78.69              & 68.38                 & 16.44                   & 4.49                   \\
		Ours                    &  &                           &                            &  & \textbf{84.96}     & \textbf{75.15}        & \textbf{10.28}                   & 2.53                   \\ \hline
		UA-MT (MICCAI'19)       &  & \multirow{7}{*}{25(10\%)} & \multirow{7}{*}{225(90\%)} &  & 80.93              & 71.31                 & 17.71                   & 5.43                   \\
		SASSNet (MICCAI'20)     &  &                           &                            &  & 79.19              & 68.80                 & 16.36                   & 6.67                   \\
		DTC (AAAI'21)           &  &                           &                            &  & 82.74              & 72.74                 & 11.76                   & 3.24                   \\
		URPC (MIA'2022 )        &  &                           &                            &  & 84.16              & 74.29                 & 11.01                   & 2.63                   \\
		SS-Net (MICCAI'22)       &  &                           &                            &  & 82.00              & 71.82                 & 10.68                   & \textbf{1.82}          \\
		MC-Net+ (MIA'2022 )     &  &                           &                            &  & 79.63              & 70.10                 & 12.28                   & 2.45                   \\
		Ours                    &  &                           &                            &  & \textbf{85.47}     & \textbf{75.97}        & \textbf{9.57}           & 2.07                   \\ \hline
		UA-MT (MICCAI'19)       &  & \multirow{7}{*}{50(20\%)} & \multirow{7}{*}{200(80\%)} &  & 85.05              & 74.51                 & 12.31                   & 3.03                   \\
		SASSNet (MICCAI'20)     &  &                           &                            &  & 82.34              & 72.72                 & 12.45                   & 3.24                   \\
		DTC (AAAI'21)           &  &                           &                            &  & 83.47              & 72.93                 & 14.48                   & 3.59                   \\
		URPC (MIA'2022 )        &  &                           &                            &  & 85.49              & 74.86                 & 8.14                    & 2.04                   \\
		SS-Net (MICCAI'22)       &  &                           &                            &  & 83.07              & 73.48                 & 10.08                   & 1.63                   \\
		MC-Net+ (MIA'2022 )     &  &                           &                            &  & 82.87              & 73.61                 & 8.94                    & 1.92          \\
		Ours                    &  &                           &                            &  & \textbf{86.31}     & \textbf{77.34}        & \textbf{7.45}           & \textbf{1.36}          \\ \hline
	\end{tabular}
\end{table*}

\begin{figure*}[!h]
	\centering
	\includegraphics[width=1.00\textwidth]{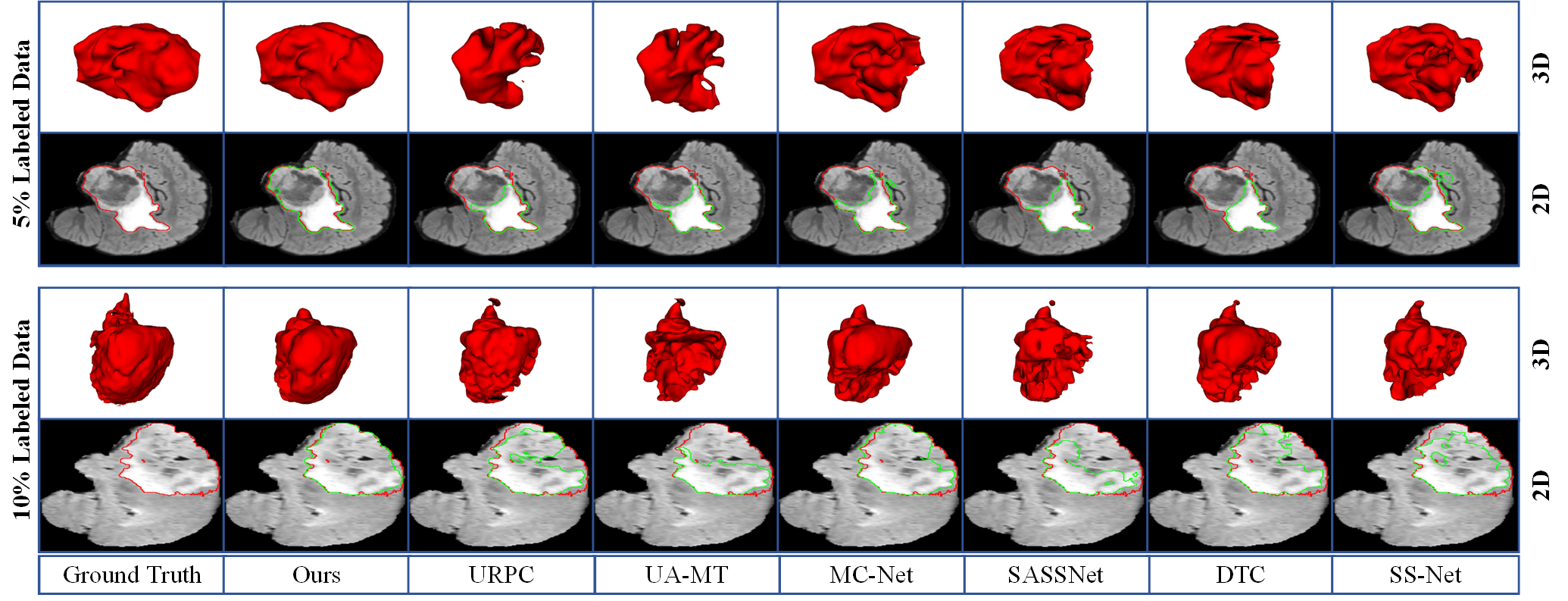}		
	\caption{2D and 3D Visualization with other methods on the BraTS2019 dataset under 5$\%$ labeled data and 10$\%$ labeled data. The red lines denote the boundary of ground truth and the green lines denote the boundary of predictions.}
	\label{figure6}
\end{figure*}

\begin{table*}[!h]
	\centering
	\caption{Ablation study about the combination of strategies on the Pancreas-CT dataset. Each Loss corresponds to a module and the Mix M. represents the Mix Module.}
	\label{table4}
	%	\resizebox{\textwidth}{!}{
	\begin{tabular}{ccccccccccccc}
		\hline
		\multicolumn{5}{c}{Designs}                                                                                                               &                      & \multicolumn{2}{c}{Scans used}                        &                      & \multicolumn{4}{c}{Metrics}                                                                   \\ \cline{1-5} \cline{7-8} \cline{10-13} 
		$L_{sup}$                 & $L_{semi}$                & Mix M.                    & $L_{consis}$              & $L_{sharp}$               &                      & Labeled                   & Unlabeled                 &                      & Dice(\%)$\uparrow$ & Jaccard(\%)$\uparrow$ & 95HD(voxel)$\downarrow$ & ASD(voxel)$\downarrow$ \\ \cline{1-5} \cline{7-8} \cline{10-13} 
		\checkmark &                           &                           &                           &                           &                      & \multirow{6}{*}{3(5\%)}   & \multirow{6}{*}{59(95\%)} &                      & 29.32              & 19.61                 & 43.67                   & 15.42                  \\
		\checkmark & \checkmark &                           &                           &                           &                      &                           &                           &                      & 48.09              & 34.59                 & 57.57                   & 25.64                  \\
		\checkmark & \checkmark & \checkmark &                           &                           &                      &                           &                           &                      & 74.01              & 59.52                 & 26.58                   & 8.83                   \\
		\checkmark & \checkmark & \checkmark & \checkmark &                           &                      &                           &                           &                      & 74.18              & 59.51                 & 8.39                    & 2.05                   \\
		\checkmark & \checkmark & \checkmark & \multicolumn{1}{l}{}      & \checkmark & \multicolumn{1}{l}{} &                           &                           & \multicolumn{1}{l}{} & \textbf{76.20}     & \textbf{62.16}        & 12.14                   & 3.49                   \\
		\checkmark & \checkmark & \checkmark & \checkmark & \checkmark &                      &                           &                           &                      & 74.69              & 60.00                 & \textbf{8.19}           & \textbf{1.74}          \\ \hline
		\checkmark &                           &                           &                           &                           &                      & \multirow{6}{*}{6(10\%)}  & \multirow{6}{*}{56(90\%)} &                      & 54.94              & 40.87                 & 47.48                   & 17.43                  \\
		\checkmark & \checkmark &                           &                           &                           &                      &                           &                           &                      & 76.40              & 63.06                 & 16.55                   & 4.75                   \\
		\checkmark & \checkmark & \checkmark &                           &                           &                      &                           &                           &                      & 80.49              & 67.84                 & 6.22                    & 1.77                   \\
		\checkmark & \checkmark & \checkmark & \checkmark &                           &                      &                           &                           &                      & 80.47              & 68.81                 & 6.16                    & 1.87                   \\
		\checkmark & \checkmark & \checkmark & \multicolumn{1}{l}{}      & \checkmark & \multicolumn{1}{l}{} &                           &                           & \multicolumn{1}{l}{} & 80.22              & 67.51                 & \textbf{5.92}           & 2.13                   \\
		\checkmark & \checkmark & \checkmark & \checkmark & \checkmark &                      &                           &                           &                      & \textbf{80.9}      & \textbf{68.4}         & 6.02                    & \textbf{1.59}          \\ \hline
		\checkmark &                           &                           &                           &                           &                      & \multirow{6}{*}{12(20\%)} & \multirow{6}{*}{50(80\%)} &                      & 71.52              & 57.68                 & 18.12                   & 5.41                   \\
		\checkmark & \checkmark &                           &                           &                           &                      &                           &                           &                      & 80.26              & 67.45                 & 15.99                   & 3.99                   \\
		\checkmark & \checkmark & \checkmark &                           &                           &                      &                           &                           &                      & 81.96              & 69.77                 & 5.18                    & 1.43                   \\
		\checkmark & \checkmark & \checkmark & \checkmark &                           &                      &                           &                           &                      & 82.51              & 70.69                 & 6.87                    & 2.11                   \\
		\checkmark & \checkmark & \checkmark & \multicolumn{1}{l}{}      & \checkmark & \multicolumn{1}{l}{} &                           &                           & \multicolumn{1}{l}{} & 81.61              & 69.35                 & 7.50                    & 2.02                   \\
		\checkmark & \checkmark & \checkmark & \checkmark & \checkmark &                      &                           &                           &                      & \textbf{82.76}     & \textbf{70.89}        & \textbf{4.85}           & \textbf{1.33}          \\ \hline
	\end{tabular}
\end{table*}
\subsubsection{Results on the LA dataset}
Table \ref{table2} presents the quantitative experimental results on the LA dataset. Compared to six SOTA methods, our model achieved optimal Dice scores across 5$\%$, 10$\%$, and 20$\%$ data annotations. With 20$\%$ labeled data, our model obtained a Dice score of 91.34$\%$, closely approaching the Dice score of 91.62$\%$ achieved through supervised learning using 100$\%$ labeled data. 
It is worth emphasizing that with 5$\%$ labeled data, we achieved a Dice score of 88.72$\%$, surpassing the accuracy of other comparative algorithms, except for MC-Net, which was trained with 10$\%$ labeled data.
Similarly, except for MC-Net, at 10$\%$ data annotation, we achieved accuracies comparable to those obtained by the other algorithms based on 20$\%$ data annotation. Simultaneously, we achieved the lowest 95HD and outstanding ASD metric.

Figure \ref{figure5} illustrates the visualization results of image segmentation for the LA dataset when trained with 5$\%$ and 10$\%$ labeled data. It is evident that our model generates a more complete and accurate left atrium than those compared models. Especially at 5$\%$ data annotation, our model naturally eliminates most of the isolated regions and preserves more fine details, whereas other models tend to produce false positive noise predictions. In the 2D view, our prediction also exhibits closer proximity to the GT. This visual representation intuitively demonstrates the effectiveness of our method.
\subsection{Results on the BraTS2019 dataset}
Table \ref{table3} presents the quantitative experimental results on the BraTS2019 dataset. In comparison to six SOTA models, our model achieved superior Dice and 95HD scores by efficiently leveraging unlabeled data across 5$\%$, 10$\%$, and 20$\%$ labeled data. Particularly, using 5$\%$ labeled data, we achieved a Dice score of 84.96$\%$, signifying a noteworthy enhancement compared to the best-scoring counterpart at 80.31$\%$ in Dice, surpassing the metric achieved by the comparative model with 10$\%$ labeled data.

Figure \ref{figure6} illustrates the visual segmentation results on the BraTS2019 dataset. Compared to six SOTA models, our model showcases a more comprehensive 3D volumetric segmentation, approaching the delineation of GT. In the 2D view, our model demonstrates closer proximity to the GT. In contrast, the compared models exhibit more false negatives relative to the somewhat ambiguous lesion features. This demonstrates the superiority of our model, showcasing its capability to identify challenging lesions with a small amount of labeled data.

\begin{figure*}[!h]
	\centering
	\includegraphics[width=0.95\textwidth]{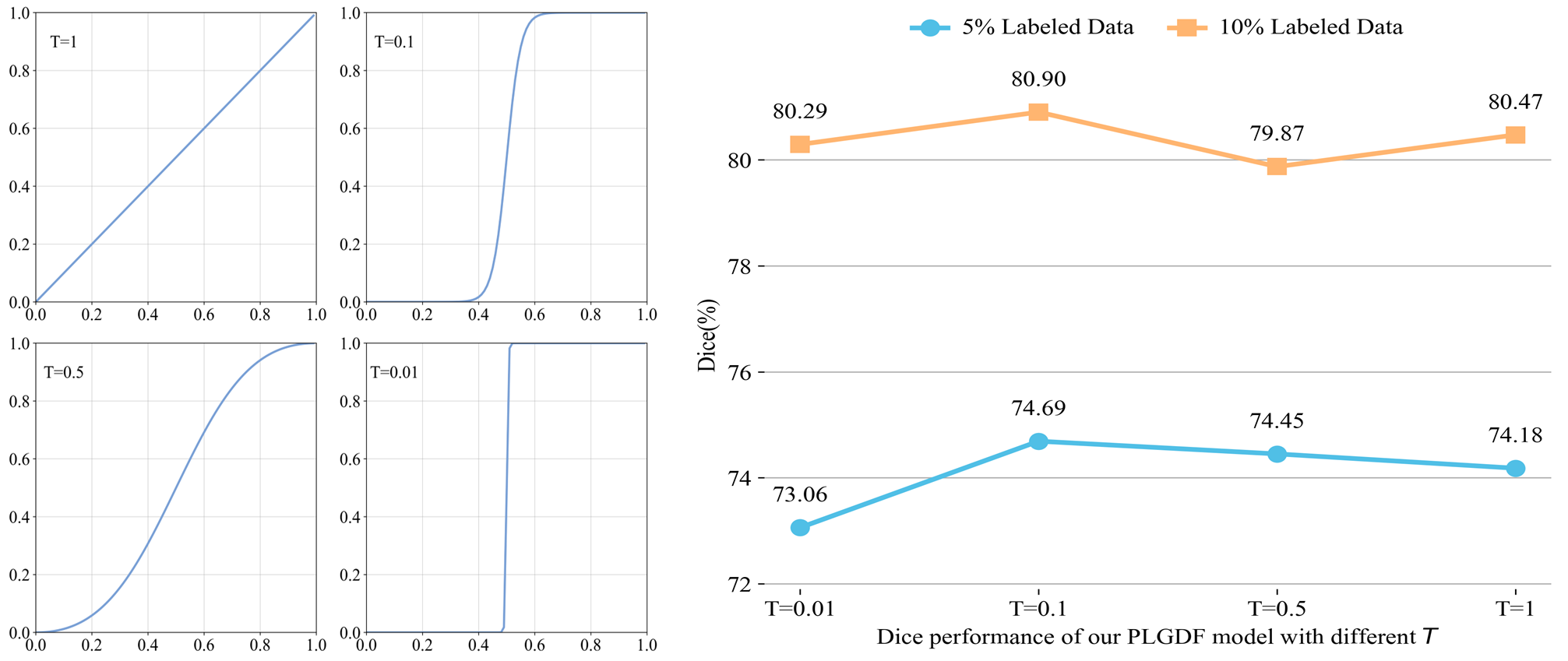}		
	\caption{Illustrations of corresponding sharpening functions (left) and dice score (right) with different sharpening temperatures T on the Pancreas-CT dataset.}
	\label{figure7}
\end{figure*}
\section{Ablation study}
In this study, we performed studies on the Pancreas-CT dataset to analyse the contributions of the effect of the contributions of each component.
\subsection{Effect of the Combination of Module.}
In our proposed framework, besides the $L_{sup}$ loss used for computing the labeled data loss, we also incorporate the Mix Module and three additional losses: $L_{semi}$, $L_{sharp}$, and $L_{consis}$, each loss represents a different module that we combined. To validate the effectiveness of these modules, we performed ablation experiments, and Table \ref{table4} presents the results obtained using different strategies.
When using only $L_{sup}$, which corresponds to utilizing a small amount of labeled data for fully supervised learning, the model's accuracy was relatively low. However, after incorporating $L_{semi}$, the model's accuracy significantly improved. Moreover, when these strategies were integrated, the results have improved steadily. This indicates that the strategies we adopted in this paper are effective and contribute to enhancing the robustness of the model.
It is worth noting that even with the combination of only the $L_{sup}$ and $L_{semi}$, the model has already achieved the best Dice and Jaccard compared to those SOTA model presented in Table \ref{table4}.
As our model incorporate $L_{consis}$ which is the strategy proposed in URPC \cite{luo2022semi}, we could spot from the table that our model also performed well without $L_{consis}$.

\subsection{Effect of the temperature $T$.}
To enhance the precision of segmentation boundaries, this study introduce a sharpening function designed to enforce entropy minimization constraints. We conducted ablation experiments on the Pancreas-CT datasets, employing 5$\%$ and 10$\%$ labeled data, aimed to assess the impact of temperature $T$ on the model's performance.
The left side of Figure \ref{figure7} delineates the application of the sharpening function across predictions for varying temperatures $T$, whereas the right side exhibits the Dice scores of the PLGDF model trained at different $T$ value on the Pancreas-CT dataset.
The outcomes demonstrate a uniformity in Dice values across different $T$ values, underscoring the model's robustness to temperature fluctuations. Elevated $T$ values are observed to insufficiently enforce entropy minimization constraints during during the training phase, while diminished $T$ values could amplify pseudo-label noise, culminating in inaccuracies. As a result, a sharpening function with a calibrated temperature of 0.1 was adopted to generate soft pseudo-labels consistently across all datasets.

\subsection{Limitations and future work}

\section{Discussion and Conclusion}
In this paper, we propose a novel semi-supervised medical image segmentation framework, named PLGDF. Building upon the mean-teacher network, we utilize the teacher model's predictions as pseudo-labels for the unlabeled data to aid in the training of the student network. Additionally, we mix the unlabeled data with labeled data to enhance dataset diversity. Furthermore, we apply a sharpening operation to the predictions of the unlabeled data to improve the clarity and accuracy of segmentation boundaries. By ensuring consistency between different scales in the decoding part of the backbone network, we further enhance the model's stability. Comprehensive experiments were performed on three publicly available medical image datasets, including the LA dataset, and BraTS2019 dataset from MR scans, and the Pancreas-CT dataset from CT scans. Remarkably, even with just 5$\%$ labeled data, our method achieves state-of-the-art results, comparable to or even surpassing the performance of state-of-the-art methods that use 10$\%$ labeled data.
These results demonstrate the effectiveness, robustness, and general applicability of our proposed framework.

Furthermore, our proposed framework is clear and combinable with some modules suitable for semi-supervised learning, such as Generative Adversarial Networks (GAN) \cite{ssl_with_gan}. While we apply sharpening operations to the predicted results, it is important to acknowledge that certain regions' predictions might already be erroneous, and sharpening these areas could potentially exacerbate misclassifications. Although the current metric suggests that sharpening brings more benefits than drawbacks, in future work, identifying correct regions and reducing interference in misjudged areas can be achieved to further improve the model's performance.
%\section{Acknowledgments}
%This work was supported by National Natural Science Foundation of China under Grant 62171133, in part by the Science and Technology Innovation Joint Fund Program of Fujian Province of China under Grant 2019Y9104, the Health Science and Technology Program of Fujian Province of China under Grant 2019-1-33, and the Industry-University-Research Cooperation Program of Fujian Province of China under Grant 2022H6006.

\section*{References}
\bibliographystyle{IEEEtran}
\bibliography{reference}

\end{document}